\documentclass[prd,twocolumn,showpacs,preprintnumbers,nofootinbib,amsmath,amssymb,nobalancelastpage]{revtex4}
\usepackage{graphicx}
\usepackage{bm}
\usepackage{dcolumn}
\usepackage{amsmath}
\usepackage{amssymb}
\usepackage{color}
\usepackage{xcolor}
\usepackage{soul}
\usepackage{url}
\usepackage{float}

\usepackage{aas_macros}
\usepackage[normalem]{ulem}

\def\H0{$H_0$}
\def\OmegaDM{\Omega_{\mathrm{DM}}}
\def\rhocrit{\rho_{\mathrm{crit}}}
\def\arec{a_{\mathrm{rec}}}

\begin{document}

\setstcolor{red}
\title{Late universe decaying dark matter can relieve the $H_0$ tension}%

\author{Kyriakos Vattis}
 \email{kyriakos_vattis@brown.edu}
\affiliation{Department of Physics, Brown University, 182 Hope St.,  Providence, RI 02912}
\affiliation{Brown Theoretical Physics Center, 182 Hope St., Providence, RI 02912}

\author{Savvas M. Koushiappas}%
 \email{koushiappas@brown.edu}
\affiliation{Department of Physics, Brown University, 182 Hope St.,  Providence, RI 02912}
\affiliation{Brown Theoretical Physics Center, 182 Hope St., Providence, RI 02912}

\author{Abraham Loeb}
\affiliation{Department of Astronomy, Harvard University, 60 Garden St., Cambridge, MA 02138}

\date{\today}

\begin{abstract}
We study the cosmological effects of two-body dark matter decays where the products of the decay include a massless and a massive particle. We show that if the massive daughter particle is slightly warm it is  possible to relieve the tension between  distance ladder measurements of the present day Hubble parameter with measurements from the cosmic microwave background. 
\end{abstract}

\maketitle

In the standard $\Lambda$CDM cosmological model, the expansion history of the universe is driven by the presence of dark matter and dark energy. Dark matter dominates the growth of structure and deceleration at early times, while dark energy is responsible for the accelerated expansion of the universe at the present epoch. Numerous experimental probes independently confirm the predictions of this model over many cosmological  scales and epochs \cite{2018arXiv180706209P}.

Despite these successes, there are hints for possible caveats to the basic assumptions of  $\Lambda$CDM. The potential discrepancies between different experimental probes seem to remain despite ongoing efforts of understanding known or, controlling the presence of, unknown systematics. 
The most prominent is the discrepancy between the value of the Hubble parameter at present time $H_0$ as it is inferred from Cosmic Microwave Background (CMB) measurements  \cite{2018arXiv180706209P} and the direct measurement in the local universe using Supenovae Type Ia (SNIa) \cite{2018ApJ...861..126R}.

The  discrepancy between the CMB measurement of \H0 and the distance ladder estimates from SNIa evolved in the last few years from $2.5\sigma$ \cite{2016ApJ...826...56R} to $3.6\sigma$ \cite{2018ApJ...861..126R}\footnote{After the initial submission of this article we learned of an improved SNIa determination of \H0 (using updated Cepheids calibration in the Large Magellanic Cloud) that increases the significance to $4.4\sigma$ \cite{2019arXiv190307603R}.}. More recently, the Dark Energy Survey (DES)  \cite{2018arXiv181102376M} found $H_0$ to be consistent with the current measurement from the CMB  \cite{2018arXiv180706209P}.  An independent measurement  of \H0 \cite{2017Natur.551...85A} was recently made possible using  gravitational waves produced from a binary neutron star merger \cite{2017PhRvL.119p1101A}. However, the uncertainty in this measurement is large (due to the single event) and cannot be used to resolve the tension. Nevertheless, future observations should be able to reduce the uncertainty to the percent level \cite{2018arXiv180705667F,2018arXiv181111723M,2018arXiv180203404F,2018arXiv180610596H,2018Natur.562..545C,2018PhRvL.121b1303V}. 

The origin of this discrepancy is unknown.  There have been suggestions that systematics are at play; however those these claims were quickly dismissed \cite{2018arXiv181002595S,2018arXiv181003526R,2018arXiv181202333V,2018arXiv181004966B}. Alternatively, attempts to relieve the tension focus on ether modifying the dark energy equation of state and its dynamics or the dark matter model. For example, a negative cosmological constant model still consistent with the data was investigated in \cite{2018arXiv180806623D} while dark energy with a dynamical equation of state was considered in \cite{2018arXiv180902340G,2019arXiv190304865K}. In \cite{2018ApJ...868...20M} the authors showed that a model with a minimally coupled and slowly or moderately rolling quintessence field cannot alleviate the discrepancy, while a more general approach was taken in \cite{2018JCAP...09..025M} where multiple models of dark energy were considered. Other proposals have been based on an early period of dark energy domination  \cite{2018arXiv181104083P} or vacuum phase transitions \cite{2018PhRvD..97d3528D,2018arXiv181011007B}.

On the other hand, modifications to the dark matter sector in order to resolve the discrepancy include partially acoustic dark matter models \cite{2017PhRvD..96j3501R}, charged dark matter with chiral photons \cite{2017PhLB..773..513K}, dissipative dark matter models \cite{2019APh...105...37D}, cannibal dark matter \cite{2018PhRvD..98h3517B} and  axions \cite{2018JCAP...11..014D}. Decaying dark matter models were also considered  in combination with solving other problems  \cite{2015PhRvD..92f1301A,2017JCAP...10..028B,2018PhRvD..98b3543B,Pandey:2019plg,1984MNRAS.211..277D,1985PAZh...11..563D,1988SvA....32..127D}.  Finally, models of interacting dark matter-dark energy \cite{2019MNRAS.482.1858Y, 2019MNRAS.482.1007Y, 2018JCAP...09..019Y} as well as modifications of the general relativity theory \cite{2018arXiv180909390E,2017arXiv171009366K,2017JCAP...10..020R} were discussed. 

Here we concentrate on the dark matter component of the cosmological model. Instead of a simple cold fluid we allow dark matter to decay into multiple components. Such a model must account for a proper treatment of the cosmological evolution of the massive decaying products \cite{2014PhRvD..90j3527B}. 

We focus on two-body decays, of the form $\psi \rightarrow \chi + \gamma$, i.e., a parent particle decays to a massless and a massive daughter particles. Below, we label the massive parent particle with a subscript $"0"$, the massless daughter with $"1"$ and the massive daughter particle with $"2"$. Such decays have been proposed in the literature in the context of Super Weakly Interacting Massive particles (Super WIMPs) \cite{2003PhRvL..91a1302F}, or decays that can explain several observables in the late universe \cite{2016PhRvD..94a5018C}. In addition, it has been recently shown that small deviations from a completely cold dark matter could be present in the late universe \cite{2018PhRvL.120v1102K}.

The proposed dark matter decays are modeled using two free parameters, namely the lifetime $\tau= 1/ \Gamma$ (where $\Gamma$ is the associated decay rate) and the fraction of rest mass energy of the parent particle transferred to the massless particle, $\epsilon$.  In this scenario, the four-momenta of the three particles involved are given by $p_{\mu,0} = ( m_0,{\bf{0}})$,  $p_{\mu,1}= ( \epsilon m_0,{\bf{p}}_1)$, $p_{\mu,2} = ( [1-\epsilon] m_0,{\bf{p}}_2)$, and the equations that govern the cosmological evolution of the massive parent and the massless daughter particles are: 
\begin{equation} 
\label{eq:rho0andrho1}
\frac{d \rho_0}{dt} + 3 \frac{\dot{a}}{a} \rho_0 = - \Gamma \rho_0 , \hspace{0.5cm} \frac{d \rho_1}{dt} + 4 \frac{\dot{a}}{a} \rho_1 = \epsilon \Gamma \rho_0, 
\end{equation} 
where $a$ is the scale factor and we assume no decays occurred prior to the redshift  of recombination $z_{\mathrm{rec}} \approx 1090$ \cite{2018arXiv180706209P}. We assign initial conditions of the dark matter density at recombination, $\rho_0(a_{\mathrm{rec}} )= \rho_{\mathrm{crit}}  \OmegaDM a_{\mathrm{rec}}^{-3}$ for any given set of $\Lambda$CDM values of $\OmegaDM$ and \H0.

The evolution of the massive daughter particle is more complex for two reasons. First this particle has a dynamic equation of state $w_2(a)$. It is possible for it to be born relativistic at some early time $a_D<1$ (when the expansion rate is given by $H_D$), but behave like matter as the universe evolves. Second, at any time the collective behavior of these particles needs to be averaged over all particles that were born prior to that interval. This means the redshift evolution of the energy density of the massive daughter particle depends on the sum of all contributions of particles born during the interval $1 \ge a \ge a_D$, some of which were born relativistic and redshifted away by $a=1$ and some that are born at late times but had no time to be redshifted). This collective behavior can be expressed as (for details, see Section II in \cite{2014PhRvD..90j3527B}) 
\begin{equation} 
\rho_2(a) = \frac{ {\cal{C}} }{a^3} \int_{a_*}^a \frac{e^{-\Gamma t(a_D)}}{a_D H_D} \left[ \frac{\beta_2^2}{1 - \beta_2^2} \left( \frac{a_D}{a} \right)^2 + 1  \right]^{1/2}   da_D, 
\label{eq:rho2} 
\end{equation}
Where $\beta_2 \equiv v_2/c =  \epsilon / ( 1 - \epsilon)$ and the constant $ {\cal{C}}$ is obtained from the initial conditions \cite{2014PhRvD..90j3527B} as  $ {\cal{C}} = \rho_{\mathrm{crit}} \OmegaDM \Gamma \exp[\Gamma t(a_{\mathrm{rec}})] \sqrt{1 - 2 \epsilon} $. 

It is important to emphasize that this approach is different from the models considered in \cite{2016JCAP...08..036P,2018PhRvD..98b3543B} where the assumption was that dark matter decays only to radiation; thus previously derived constraints and conclusions do not apply here.

We can understand qualitatively the effects of a decay in this scenario in the following way. For a fixed dark matter density $\OmegaDM$ and lifetime $\tau$, increasing $\epsilon$ lowers the value of $H(z=0)$ at low redshifts as more non-relativistic energy density is transferred to radiation whose energy density is diluted  at a faster rate as the universe expands.  On the other hand, keeping the value of $\epsilon$ constant and decreasing the lifetime of the parent particle $\tau$ shifts the  matter-dark energy equality to earlier times and also decreases the value of $H(z=0)$ as more dark matter will decay by the present epoch. Therefore  a combination of $\epsilon$ and $\tau$ could bring the measured value of the expansion rate at $z=0$ in agreement with the evolution of $H(z)$ at higher redshifts as measured at recombination.

\begin{figure}
\includegraphics[scale = 0.35]{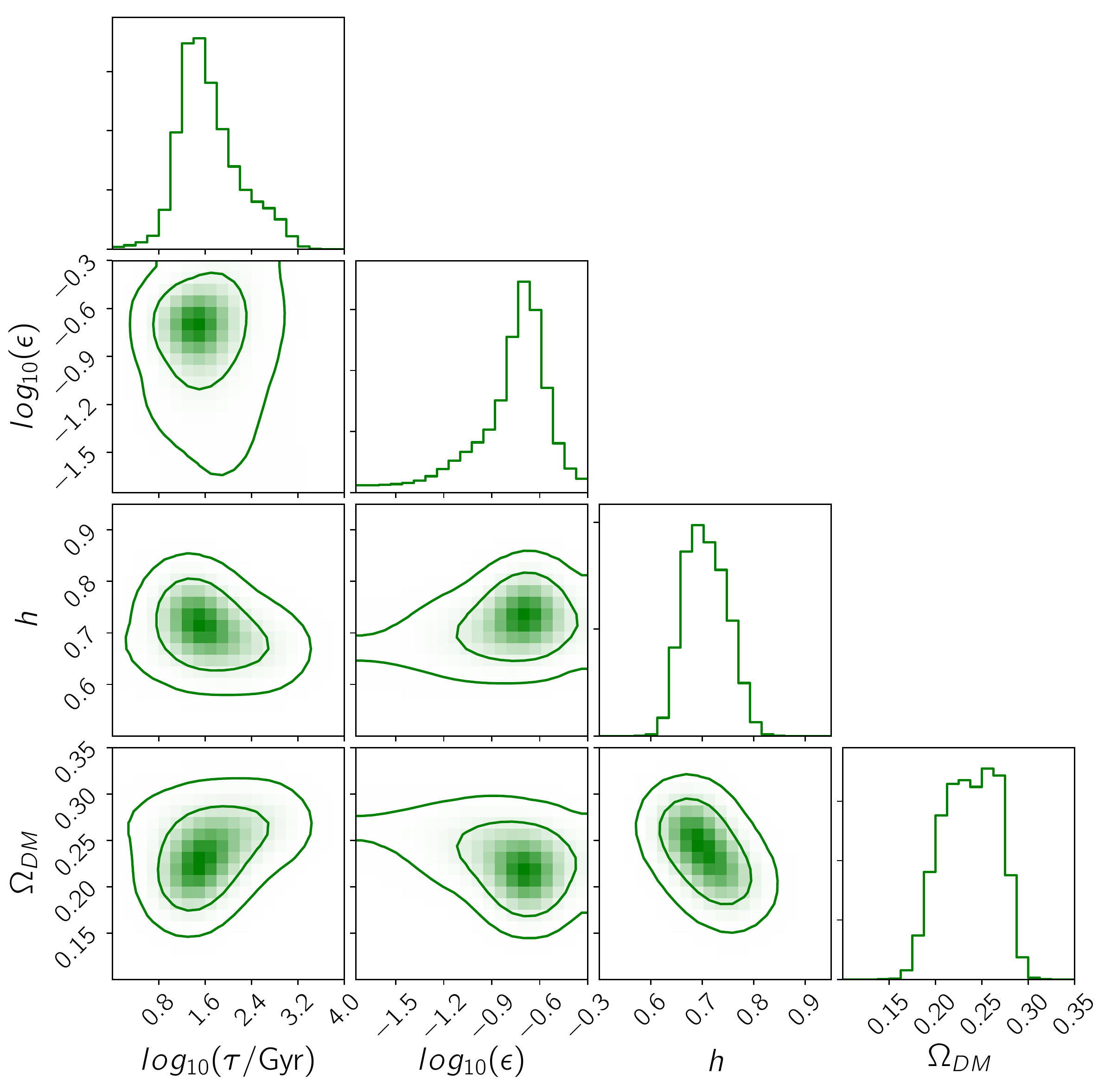}
\caption{\label{fig:figure1}  Results of the MCMC analysis of the decaying dark matter scenario with energy fraction $\epsilon$ and decay time $\tau$ using the late universe data together with effective data points between z=3 and 1090. The Hubble parameter $h$ and matter density $\OmegaDM$ represent the values used in the $ \Lambda$CDM universe that sets the initial conditions for the decaying dark matter model and not the values obtained in the latter (see text for details).
}
\end{figure}

Given the above considerations, we use  the Friedmann equation for a flat geometry 
\begin{equation}
\label{eq:Hk}
H^2(a) \equiv \left(   \frac{\dot{a}}{a}  \right)^2 = \frac{8 \pi G}{3}  \sum_i \rho_i(a) , 
\end{equation} 
where 
\begin{eqnarray} 
\sum_i \rho_i(a) &=& \rho_0(a) + \rho_1(a) + \rho_2(a) \nonumber \\
&+& \rho_r(a) + \rho_\nu(a) + \rho_b(a) + \rho_\Lambda, 
\end{eqnarray} 
and perform a Monte Carlo Markov Chain (MCMC) analysis of a decaying dark matter cosmology under the assumption that no decays have taken place prior to recombination (in other words, the universe at recombination is correctly described by CMB measurements \cite{2018arXiv180706209P}, and $\tau \gg 400,000$ years).

\begin{figure}
\includegraphics[scale=0.35]{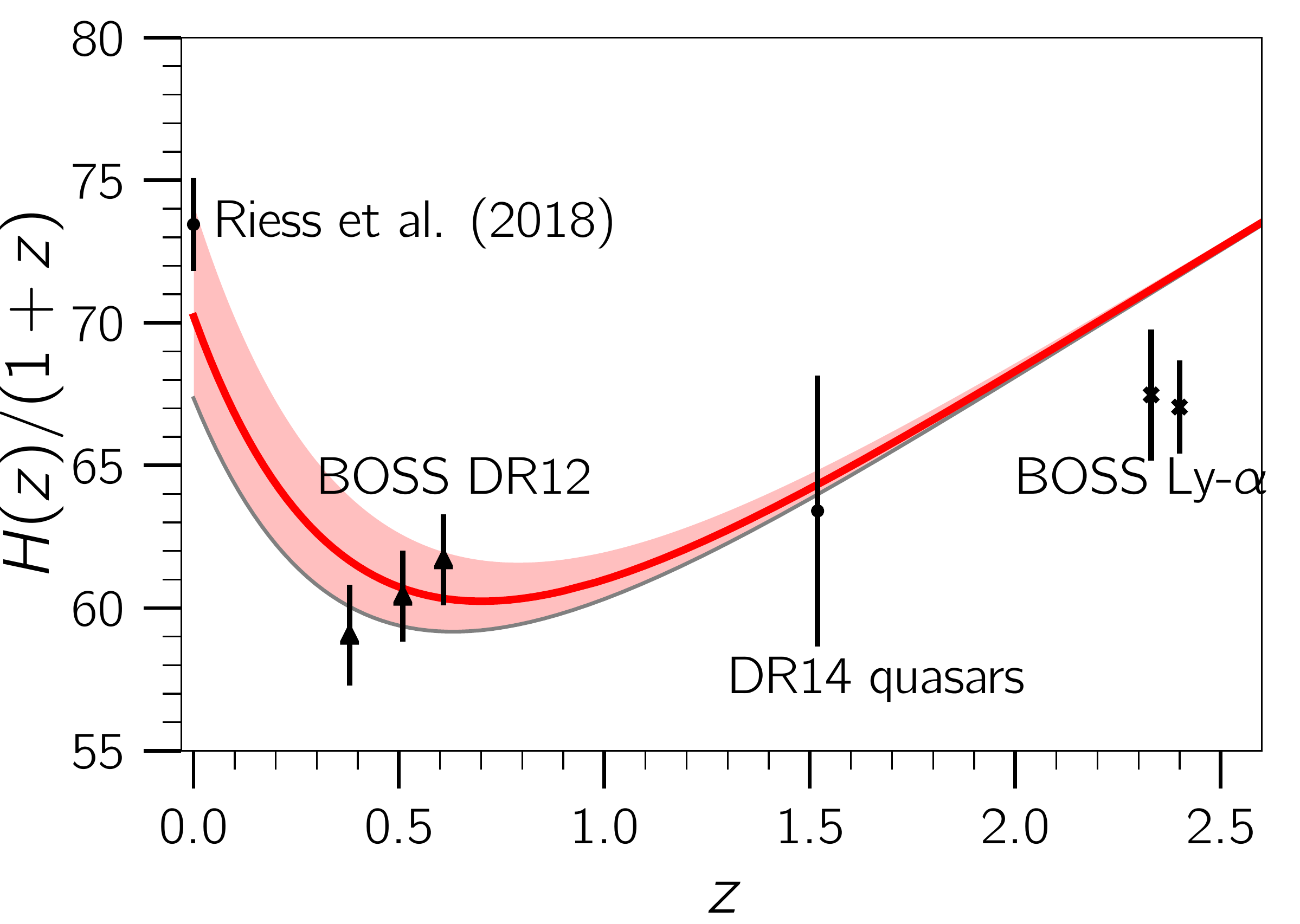} 
\caption{The evolution of the Hubble parameter as a function of redshift for $\Lambda$CDM (thin grey line) and the proposed decaying dark matter scenario. Thick red line depicts the median while the red shaded area represents the $68\%$ allowed interval. The units of the y-axis are $\rm{km}\rm{s}^{-1}\rm{Mpc}^{-1}$.   The decaying dark matter scenario proposed here can ease the tension between Planck18   \cite{2018arXiv180706209P} and the SHOES measurement of $H_0$ \cite{2018ApJ...861..126R} while  matching Planck18 $\Lambda$CDM universe at high redshifts. }
\label{fig:figure2}
\end{figure}

\begin{figure*}
\includegraphics[scale=0.35]{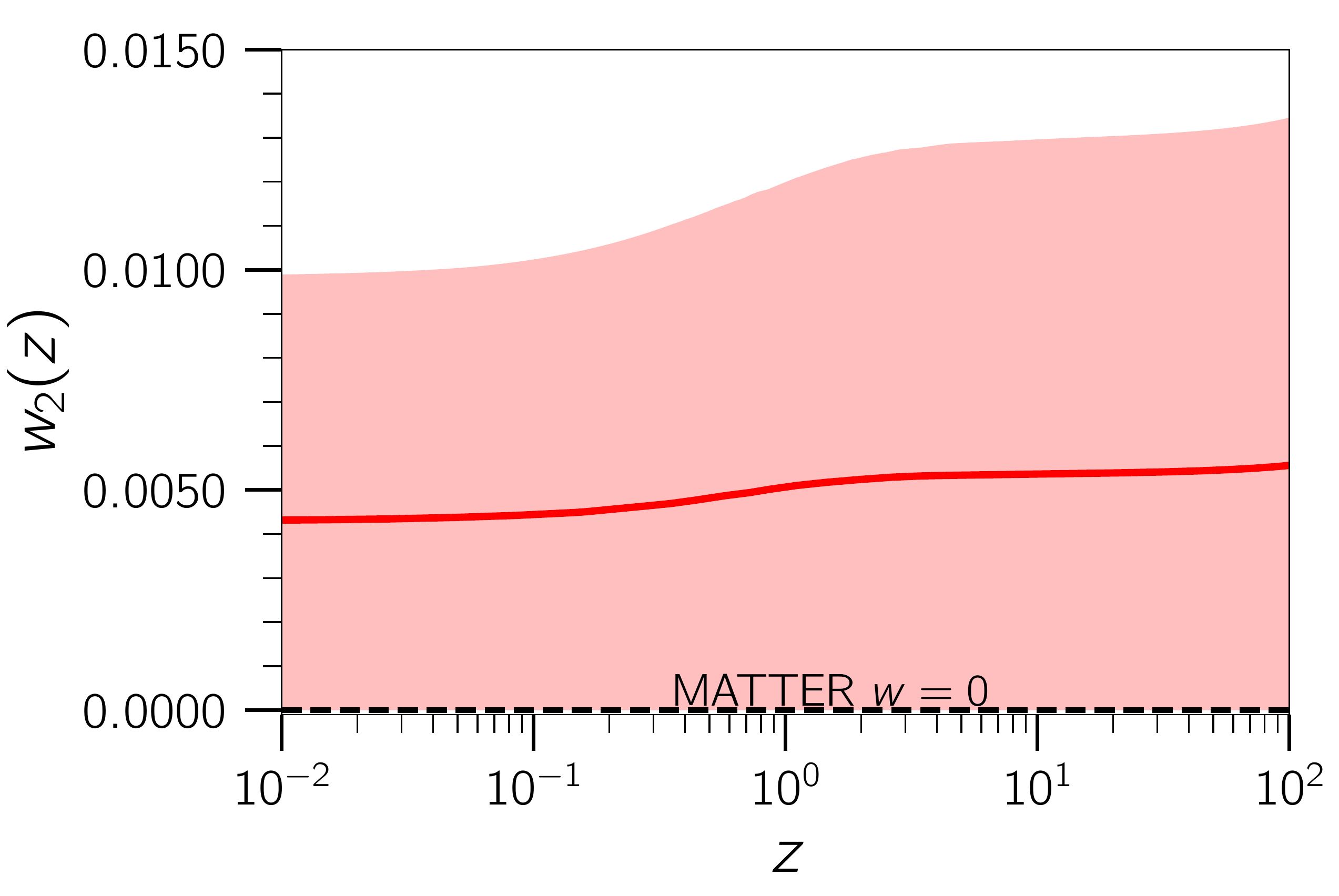}\includegraphics[scale=0.35]{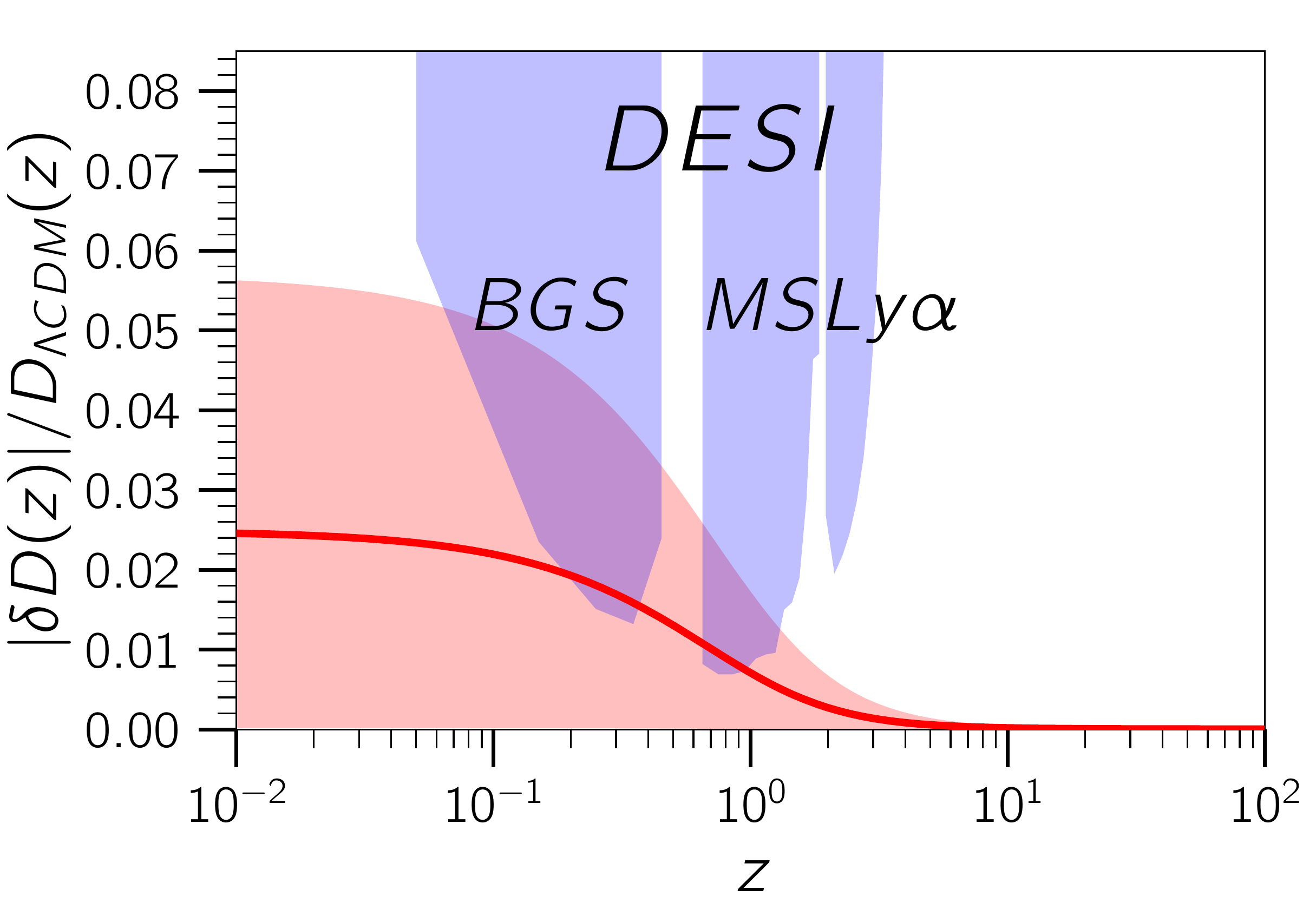}
\caption{\label{fig:manybodyD} {\it{Left}}:  The evolution of the equation of state $w_2$ of the massive daughter particle as a function of redshift. The shaded area represents the 68 percentile region. {\it{Right}}: Ratio of the difference for the linear growth factor between the decaying dark matter scenario proposed here and $\Lambda$CDM. Future surveys such as  the Dark Energy Spectroscopic Instrument (DESI) \cite{2016arXiv161100036D} will be able to test the proposed decaying dark matter model. At lower redshifts DESI will constrain the growth factor with the Bright Galaxy Survey (BGS), around $z\approx 1$ , with the Main Survey (MS) and at higher redshifts with the Lyman $\alpha$ survey (Ly-$\alpha$) \cite{2016arXiv161100036D}. }
\label{fig:figure3}
\end{figure*}

We allow four free parameters, $\tau$, $\epsilon$, $\OmegaDM$, and  $h = H_0 / (100 {\mathrm{km/s \, Mpc^{-1}}})$. We assume flat logarithmic priors for the dark matter decaying parameters in the ranges $-4 \le \log_{10} \epsilon < \log_{10}(0.5)$, $ -3 \le \log_{10}\tau \le 4$, and flat priors in the ranges $0\le \OmegaDM \le 1$ and $0.5\le h \le 1$ for the matter density and the hubble parameter. A sample of $\OmegaDM$, and $h$ (thus $\rhocrit$) give the initial conditions for the dark matter energy density at recombination (scale factor $\arec$) as $\rho_0 = \OmegaDM \rhocrit \arec^{-3}$ and $\rho_1 = \rho_2 = 0$ (taking $\arec=1/(1+z_{\mathrm{rec}})$, where $z_{\mathrm{rec}} \approx 1090$ \cite{2018arXiv180706209P}. Note that the sampled choice of $h$ (in setting the initial conditions) is not the same as the value of $H(z=0)$ derived {\it{after}} evolving the universe from recombination to $z=0$. 
For the remainder of the species we use the Planck TT,TE,EE+lowE+lensing model to obtain their densities at recombination with the exception of dark energy which is chosen by enforcing flatness. To take into account the neutrino mass we follow section 3.3 of Ref.~\cite{2011ApJS..192...18K}. 

The aforementioned procedure guarantees the universe behaves like $\Lambda$CDM  at recombination and sets the initial conditions from which we then solve equation~(\ref{eq:Hk}) (with  equations~\ref{eq:rho0andrho1} \& \ref{eq:rho2}) from $a=\arec$ to $a=1$. 
We run the MCMC analysis against the following late universe measurements of $H(z)$: the distance-ladder Hubble measurement \cite{2018ApJ...861..126R}, the BAO measurements from BOSS DR12 \cite{2017MNRAS.470.2617A}, from BOSS DR14 quasars \cite{2018MNRAS.477.1639Z}, BOSS Ly-$\alpha$ auto-correlation at z = 2.33 \cite{2017A&A...603A..12B} and the joint constraint from the Ly-$\alpha$ auto-correlation and cross-correlation with effective redshift z = 2.4 \cite{2017A&A...608A.130D}. 

The result of this analysis is shown in Figure~\ref{fig:figure1}. The inner and outer contours correspond to the 68 and 96 percentiles of the projected two-dimensional space of each panel. As expected, $h$ and $\OmegaDM$ are anti-correlated, in agreement with the earlier analysis of \cite{2016PhRvD..93b3510B}. This is to be expected as increasing $h$ requires the universe to expand faster at late times which means the matter-dark energy equality must move to earlier times which necessitates a lower value of $\OmegaDM$. This effect can also be seen as the correlation between the lifetime of the particle $\tau$,  $\OmegaDM$ and $h$ --  for smaller lifetimes and larger fraction energy transferred to radiation, $h$ and $\OmegaDM$ need to adjust accordingly by increasing and lowering their values respectively to maintain agreement with the data. In addition, for lower values of $\tau$ higher values of $\epsilon$ are required,  implying that more energy is needed to be transferred from matter to radiation and thus the deviation from  $\Lambda$CDM occurs over a shorter period of time. Note that $h$ and $\OmegaDM$ in Figure~\ref{fig:figure1} are not the derived values of the Hubble expansion rate and matter density obtained from the evolved decaying dark matter scenario but the sampled values that are used to set the initial conditions (i.e., the values used to obtain the dark matter density at recombination).

The elongation of the 96 percentile contour of $\epsilon$ is due to the fact the data used seems to be well fitted by Planck's $\Lambda$CDM model with the only exception of the distance-ladder measurement of the Hubble parameter \cite{2018ApJ...861..126R} and therefore this extension to very small values of $\epsilon$ encapsulates the tendency of the allowed parameter space to approach $\Lambda$CDM (a universe with $\epsilon=0$ is identical to $\Lambda$CDM). 


In figure~\ref{fig:figure2} we show the redshift evolution of the Hubble parameter. At early times (prior to the onset of decays) the universe behaves identically to the $\Lambda$CDM universe inferred from CMB measurements. At late times, decays (and the resultant transfer of energy from matter to radiation) speed up the expansion and results in a late universe measurement of $h$ that is higher than the one obtained  from the CMB under $\Lambda$CDM and thus potentially alleviating the tension. 

The derived 68 confidence limits for each one of the free parameters are shown in Table~\ref{tab:table1} which is the main result of this paper -- a 2 body decay with a relatively high value of $\epsilon$ and a long lifetime (significantly longer than the age of the universe) can relieve the tension between measurements of the present expansion rate measured from the CMB and the local universe. 
The model works because it allows for a fraction of the rest mass energy of the parent particle to go into radiation, with the remainder going to the speed of the massive daughter particle.  The left panel in figure~\ref{fig:figure2} shows the evolution of the equation-of-state of the massive daughter particle as a function of redshift, which shows that it evolves as $\rho_2 (a) \propto a^{-3.15}$, as expected from a particle that is not fully relativistic or fully cold. 

Note that part of the allowed parameter space (lifetimes less than $\le 40$ Gyrs) has already been constrained by numerical simulations of decaying dark matter together with the Sloan Digital Sky Survey (SDSS) Lyman$-\alpha$ power spectrum \cite{PhysRevD.88.123515}. These constraints are obtained by comparing dark matter-only simulations together with the fluctuation Gunn-Peterson approximation \cite{1998MNRAS.296...44G,1998MNRAS.296...44G} against the SDSS Lyman$-\alpha$ power spectrum \cite{McDonald_2006,2013A&A...559A..85P} (see however the difficulties in understanding the systematics relevant to damped Lyman$-\alpha$ systems \cite{Alonso_2018}). However, the analysis presented here is not simply an estimate of the lifetime of the dark matter parent particle, but a combination of the four free parameters that can alleviate the \H0 tension -- in other words the preferred value of the lifetime of the particle may not correspond to the largest change in the value of \H0. 

The allowed parameter space encompasses dark matter decays that can alleviate some of the small scales problems that exist in the context of galaxy formation \cite{2014MNRAS.445..614W}, as well as the amplitude of linear fluctuations \cite{2015JCAP...09..067E}. For example, a dark matter particle with a long lifetime ($\sim150$ Gyrs), and a value of $\epsilon \sim 10^{-2}$ provides a boost of $\gamma_2 = [1 - \epsilon^2 / ( 1 - \epsilon)^2]^{-1/2} \approx 1.00005$, a characteristic velocity dispersion of order ${\cal{O}} (10^3){\mathrm{km/s}}$ and a fraction of dark matter that has already decayed of  few percent. 

\begin{table}[tbp]
\centering
\caption{$68\%$  Confidence Limits}
\label{tab:table1}
\begin{tabular}{ccccc}
\hline
$\log_{10}\epsilon$ 	& 	$\log_{10}(\tau/\mathrm{Gyr}])$ 	& 	$\OmegaDM$ 	&	$h$ \\
\hline
$-0.78^{+0.14}_{-2.10} $ &     $ 1.55^{+0.63}_{-0.25}$ & $0.24^{+0.03}_{-0.03}$ & $0.70^{+0.04}_{-0.03}$\\
\hline
\end{tabular}
\end{table}

Dark matter decays will also affect the growth of structure. We can quantify the effects of decays on structure by calculating the linear growth factor $D(z)$ as the growing mode solution to the differential equation that governs the linear evolution of matter perturbations,
\begin{equation}
\label{eq:D}
\frac{d^2D}{da^2}+\left(\frac{d \ln H}{da}+\frac{3}{a}\right)\frac{dD}{da}-\frac{4\pi G\rho_m}{a^2}=0,
\end{equation}
where $\rho_m=\rho_0+\rho_2+\rho_b$ and D is normalized to unity today. Here, we assume that the massive daughter contributes to the matter content of the universe; in reality (as we show above) this is not entirely correct as the particle is warm, and therefore the derived result is a conservative upper bound to the effect. 
The right panel in figure~\ref{fig:figure3} shows the deviation of the linear growth factor from the standard $\Lambda$ cosmology. Future surveys, such as DESI  \cite{2016arXiv161100036D} will be able to test the predictions of this decaying dark matter scenario (see also \cite{PhysRevD.85.043514}).

Finally, we can put the derived constraints on $\tau$ and $\epsilon$ in the context of particle physics models that include Super WIMPs or exited dark fermions that can decay to a lighter fermion and a photon via a magnetic dipole transition \cite{2003PhRvL..91a1302F,2016PhRvD..94a5018C}. In general, the rate will be given by $\Gamma \sim \delta m^3 / \Lambda^2$ where $\Lambda$ is some high scale. The mass difference can be obtained from the kinematics of this two-point decay and is given by $\delta m = 1 - \sqrt{(1 - 2 \epsilon)}$ in units of the parent particle. Therefore a lower limit of $\epsilon \approx 0.17$  and a lifetime of the parent particle of $\tau \approx 20 {\mathrm{Gyr}}$ implies that for a ${\mathrm{GeV}}$ scale  particle $\delta m \approx 180 {\mathrm{MeV}}$ with $\Lambda \approx 10^{16} \mathrm{GeV}$.

In summary, we have shown that dark matter decays of the form of $\psi \rightarrow \chi + \gamma$ potentially can relieve the \H0 tension between the value obtained by the local measurements using the distance ladder and the value obtained by observations of the CMB. Further  analysis of the related effects on large scale structure formation and the CMB power spectrum is required to confirm to extent at which decays can solve the \H0 problem. \\

\acknowledgments
We acknowledge useful conversations with Manuel Buen$-$Abad, Jatan Busch, Ian Dell'Antonio, JiJi Fan, John Leung, David Pinner, Vivian Poulin, Robert Sims, Michael Turner and Andrew Zentner. KV and SMK are supported by DE-SC0017993. AL was  supported in part by a John Templeton Foundation grant awarded to the Black Hole Initiative at Harvard University. 

\bibliography{manuscript}

\end{document}